\documentstyle[preprint,aps,epsfig]{revtex}

\tightenlines
\preprint{\vbox{  
\hbox{IFT-P.026/2000}   
\hbox{March 2000} }}  
\begin{document}  
\draft 
\title{
Left-right asymmetries and exotic vector--boson discovery in lepton-lepton
colliders}
\author{J. C. Montero, V. Pleitez and M. C. Rodriguez} 
\address{
Instituto de F\'\i sica  Te\'orica\\ 
Universidade  Estadual Paulista\\
Rua Pamplona, 145\\ 
01405-900-- S\~ao Paulo, SP\\ 
Brazil } 
\maketitle 
\begin{abstract}  
By considering  left-right (L-R)
asymmetries we study the capabilities of lepton colliders in searching 
for new exotic vector bosons. Specifically we study the effect of a 
doubly charged bilepton boson and an extra neutral vector boson appearing 
in a 3-3-1 model on the L-R asymmetries for the processes 
$e^-e^-\to e^-e^-$, $\mu^-\mu^-\to\mu^-\mu^-$ and
$e^-\mu^-\to e^-\mu^-$ and show that these asymmetries are very 
sensitive to these new contributions and that they are in fact  powerful tools 
for discovery this sort of vector bosons.    

\end{abstract}

\section{Introduction}
\label{sec:intro}
Any extension of the electroweak standard model (ESM)~\cite{sm} implies 
necessarily the existence of new particles. We can have a rich scalar-boson 
sector if there are several Higgs-boson multiplets~\cite{higgshunter} or have 
more vector and scalar fields in models with a larger gauge symmetry as in 
the left--right symmetric~\cite{lr} and in 3-3-1 models~\cite{331}, or we 
also  can have at the same time more scalar, fermion, and vector particles 
as in the supersymmetric extensions of the ESM~\cite{mssm}.

If in a given model all the new particles contribute to all observables, it
will be very difficult to identify their contribution in the usual and exotic
processes. In some models~\cite{331,derli} the contributions of the
scalar-bosons can not be suppressed by the fermion mass and they can have same
strength of the fermion--vector-boson coupling. Hence, we can ask ourselves if
there exist observables and/or processes which allow us to distinguish between 
the contributions of charged and neutral scalar-bosons from those of the
vector-bosons. In Ref.~\cite{assi1} it was noted that the left-right 
(L-R) asymmetries in the lepton--lepton diagonal scattering are insensible 
to the contribution of doubly-charged scalar fields but are quite sensible to 
doubly-charged vector field contributions. On the other hand, in non-diagonal 
scattering (as $\mu^-e^-$) those asymmetries are sensible to the existence 
of an extra neutral vector-boson $Z^\prime$~\cite{assi2,langacker}. 

Here we will extend our previous analysis by considering a detailed study of
the L-R asymmetries in order to analyse their capabilities in detecting new
physics. The outline of the paper is the following: In Sec. II we define the
asymmetries; in Sec. III we show the lagrangian interaction of the models we
are considering here. The results and experimental considerations are given 
in Sec. IV and our conclusions appear in the last section. 

\section{The L-R asymmetries}
The left-right asymmetry for the process $l^-l^{\prime-} \to l^-l^{\prime-}$ 
with one of the particles being unpolarized is defined as
\begin{equation}
A_{RL}(ll^{\prime} \to ll^{\prime})\equiv A_{RL}(ll^{\prime})=
\frac{d\sigma_R-d\sigma_L}{d\sigma_R+d\sigma_L},
\label{asy1}
\end{equation}
where $d\sigma_{R(L)}$ is the differential cross section for one right
(left)-handed lepton $l$ scattering on an unpolarized lepton $l^{\prime}$ and
where $l,l^{\prime}=e,\mu$. 
That is
\begin{equation}
 A_{RL}(ll^{\prime})=
\frac{(d\sigma_{RR}+d\sigma_{RL})-(d\sigma_{LL}
+d\sigma_{LR})}{(d\sigma_{RR}+d\sigma_{RL})+(d\sigma_{LL}
+d\sigma_{LR})},
\label{asy2}
\end{equation}
where $d\sigma_{ij}$ denotes the cross section for incoming leptons with
helicity $i$ and $j$, respectively, and they are given by 
\begin{equation}
d\sigma_{ij}
\propto\sum_{kl}\vert M_{ij;kl}\vert^2,\quad i,j;k,l=L,R.
\label{dsigma}
\end{equation}
Notice that
when the scattering is diagonal, $l=l^{\prime}=e,\mu$,  
$d\sigma_{RL}=d\sigma_{LR}$,
so the asymmetry in Eq.~(\ref{asy2}) is equal to the asymmetry defined as
$(d\sigma_{RR}-d\sigma_{LL})/(d\sigma_{RR}+d\sigma_{LL})$. 
For practical purposes, for the non-diagonal ($e\mu\to e\mu$) case of the 
$A_{RL}$ asymmetry, we will focus on the scattering of polarized muons by 
unpolarized electrons.   

Another interesting possibility is the case when both leptons are 
polarized. We can define an asymmetry $A_{R;RL}$ in which
one beam is always in the same polarization state, say right-handed, and 
the other  is either
right- or left-handed polarized (similarly we can define  $A_{L;LR}$):
 \begin{equation}
A_{R;RL}=\frac{d\sigma_{RR}-d\sigma_{RL}}{d\sigma_{RR}+d\sigma_{RL}},\qquad
A_{L;RL}=\frac{d\sigma_{LR}-d\sigma_{LL}}{d\sigma_{LL}+d\sigma_{LR}}.\qquad
\label{ar}
\end{equation}
In this case, when the non-diagonal scattering is considered, we will assume 
that the muon beam has always the same polarization and the electron one can 
have both, the left and the right polarizations.

We can integrate over the scattering angle and define the asymmetry 
$\int A_{RL}$ as

\begin{equation}
\int A_{RL}=\frac{(\int d\sigma_{RR}+\int d\sigma_{RL})-(\int 
d\sigma_{LL}
+\int d\sigma_{LR})}{(\int d\sigma_{RR}+\int d\sigma_{RL})+(\int d\sigma_{LL}
+\int d\sigma_{LR})},
\label{aint}
\end{equation}

where $\int d\sigma_{ij}\equiv \int^{175^o}_{5^o}d\sigma_{ij}$. A similar
expression can be written for $A_{R;RL}$.

\section{The models}
\label{sec:models}

We are studying here the asymmetries defined above in the context 
of two models: the electroweak standard model (ESM) and in a model having 
a doubly charged bilepton vector field ($U^{--}_\mu$) 
and an extra neutral  vector boson $Z^\prime$~\cite{331}. 
The latter model has also two doubly charged scalar bileptons but since their 
contributions cancel out in the numerator of the asymmetries we are not 
consider them on this study. 
We identify the case under study by using the $\{{\rm ESM}\}$, 
$\{{\rm ESM+U}\}$, and $\{{\rm ESM+Z^\prime}\}$ labels in cross sections 
and asymmetries.

In the context of 
the electroweak standard model, at the tree level, the relevant part of the 
ESM lagrangian  is
\begin{equation}
{\cal L}_F=-\sum_i\,\frac{g\,m_i}{2M_W}\,\bar\psi_i\psi_i\,H^0-
e\sum_iq_i\bar\psi_i\gamma^\mu\psi_iA_\mu-\frac{g}{2\cos\theta_W}\,
\psi_i\gamma^\mu(g^i_V-g^i_A\gamma^5)\psi_iZ_\mu,
\label{su21}
\end{equation}
$\theta_W\equiv\tan^{-1}(g'/g)$ is the weak mixing angle, $e=g\sin\theta_W$
is the positron electric charge with $g$ such that
\begin{equation}
g^2=\frac{8G_F M^2_W}{\sqrt2};\quad {\rm or}\quad 
g^2/\alpha=4\pi\sin^2\theta_W,
\label{gf}
\end{equation}
with $\alpha\approx 1/128$; and the vector and axial neutral couplings are
\begin{equation}
g^i_V\equiv t_{3L}(i)-2q_i\sin^2\theta_W,\quad
g^i_A\equiv t_{3L}(i),
\label{gva}
\end{equation}
where $t_{3L}(i)$ is the weak isospin of the fermion $i$ and $q_i$ is the 
charge of $\psi_i$ in units of $e$. 

The charged current interactions in a model having a doubly charged vector 
boson\cite{331}, in terms of the 
physical basis, are given by
\begin{equation}
-\frac{g}{\sqrt2}\left[\bar\nu_L E^{\nu\dagger}_LE^l_Ll_LW^+_\mu+ 
\bar l^c_L\gamma^\mu E^{lT}_RE^\nu_L \nu_L V^+_\mu-
\bar l^c_L E^{lT}_RE^l_L l_L U^{++}_\mu\right]+H.c.,
\label{cc331}
\end{equation}
with $l'_L=E^l_Ll_L,\quad l'_R=E^l_Rl_R,\quad \nu'_L=E^\nu_L\nu_L$,
the primed (unprimed) fields denoting symmetry (mass) eigenstates.
We see from Eq.~(\ref{cc331}) that for massless neutrinos we have no mixing
in the charged current coupled to $W^+_\mu$ but we still have mixing 
in the charged currents coupled to $V^+_\mu$ and $U^{++}_\mu$. 
That is, if neutrinos are massless we can always chose 
$E^{\nu\dagger}_LE^l_L=1$. However,  
the charged currents coupled to $V^+_\mu$ and $U^{++}_\mu$ are not diagonal in 
flavor space and the mixing matrix ${K}=E^{lT}_RE^\nu_L$ has
three angles and three phases. (An arbitrary $3\times 3$ unitary matrix has 
three angles and six phases. In the present case, however, the matrix 
${\cal K}$ is determined entirely by the charged lepton sector, so we 
can rotate only three phases~\cite{liung}).

The total width of the $U$-boson ($\Gamma^{\mbox{total}}_U$) is a calculable 
quantity in the model once we know all the $U$-boson couplings which are 
derived from the 3-3-1 gauge-invariant lagrangian. However, we find that a 
complete computation of $\Gamma_U$ is out of the scope of this paper because 
in this case some realistic hypotheses  concerning the masses of the 
exotic scalars and quarks should be made. Thus, we will only consider the
partial width due to the $U^{--} \to l^-l^-$ decay. In the limit where all the
lepton masses are negligible we have:
\begin{equation}
\Gamma^{\mbox{total}}_U\sim \Gamma(U^{--} \to \mbox{leptons} ) = 
\sum_{i,j} \frac{G_F}{6\sqrt2 \pi}M_W^2 M_U \vert
K_{ij}\vert^2
\label{largu}
\end{equation}
where $i,j$ run over the $e,\mu$, and $\tau$ leptons and $K_{ij}$ is a mixing
matrix in the flavor space. For the expression above we can write
$\Gamma^{\mbox{total}}_U = \sum_i \Gamma_{ii}+
\frac{1}{2}\sum_{i\not=j}\Gamma_{ij}$ and assuming that the matrix $K$ is
almost diagonal we can neglect $\Gamma_{ij}$ for $i\not= j$ and consider for
practical purposes that  $\Gamma^{\mbox{total}}_U = 3\times \Gamma_{ii}$. 
In our numerical applications $\Gamma^{\mbox{total}}_U$ is a varying function 
of the $U$-boson mass. For instance, 
for $M_U=300$ GeV we have $\Gamma^{\mbox{total}}_U \sim 2.5$ GeV.

In the model there is also a $Z'$ neutral vector boson which couples with the
leptons as follows
\begin{equation}
{\cal L}_{NC}^{Z'}=-\frac{g}{2c_W}\,\left[\bar{l}_{aL}\gamma^\mu L_l{l}_{aL}
+\bar{l}_{aR}\gamma^\mu R_l{l}_{aR}+\bar{\nu}_{aL}\gamma^\mu L_\nu{\nu}_{aL}
\right]\,Z'_\mu,
\label{zpnc}
\end{equation}
with $L_l=L_\nu=-(1-4s^2_W)^{1/2}/\sqrt3$ and $R_l=2L_l$. Notice the 
{\it leptophobic} character of  $Z^\prime$\cite{dumm}. In this case we have
no concerns about the $Z^\prime$--width because this neutral boson is only
exchanged in the $t$--channel.

We will consider the process
\begin{equation}
l^-(p_1,\lambda)+l^{\prime-}(q_1,\Lambda)\to 
l^-(p_2,\lambda')+l^{\prime-}(q_2,\Lambda'),
\label{eumu}
\end{equation}
where $q=p_2-p_1=q_2-q_1$ is the transferred momentum. As we said before, 
we will neglect the electron mass but not the muon mass {\it i.e.}, 
$E=\vert\vec{p}_e\vert$ for the electron and 
$K^2-\vert\vec{q}_\mu\vert^2=m_\mu^2$ for the muon. In the non-diagonal 
elastic scattering in the standard model we have only the $t$-channel 
contribution. The relevant amplitudes for the ESM,
$\{{\rm ESM+U}\}$ and $\{{\rm ESM+Z^\prime}\}$ models 
are in  the appendices of Ref.~\cite{assi1} (Ref.~\cite{assi2}) for the 
diagonal (non-diagonal) case.

\section{Results}
\subsection{The $U$ boson}

\begin{figure}[ht]
\begin{center}
\vglue -0.009cm
\mbox{\epsfig{file=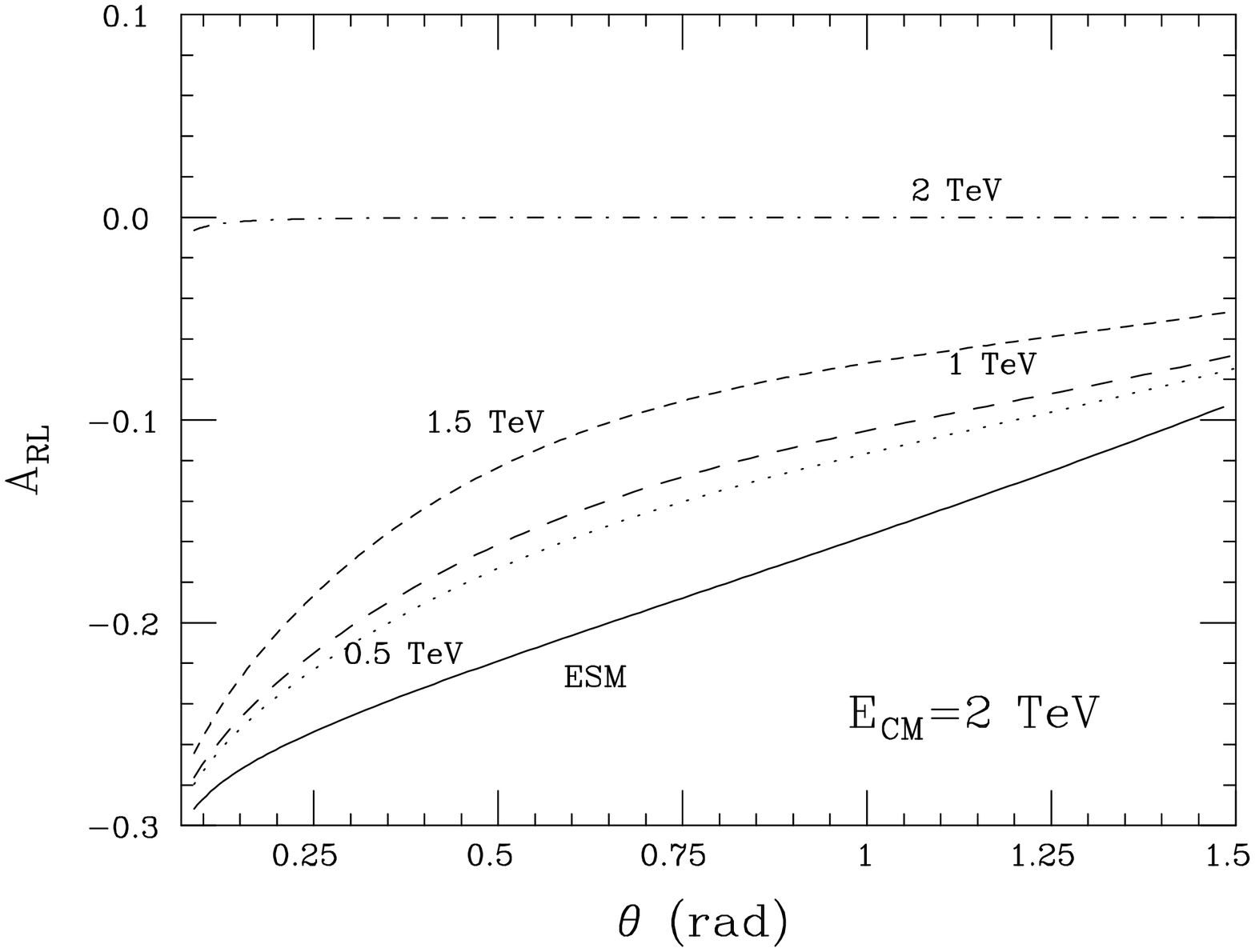,width=0.7\textwidth,angle=90}}       
\end{center}
\vglue 2cm
\caption{The $A_{RL}$ asymmetry for an $e^-e^-$
collider with $\sqrt{s}=2$ TeV for the ESM (solid line) and for the ESM+U for
several U--masses as a function of the scattering angle.}
\label{fig1}
\end{figure}

\begin{figure}[ht]
\begin{center}
\vglue -0.009cm
\mbox{\epsfig{file=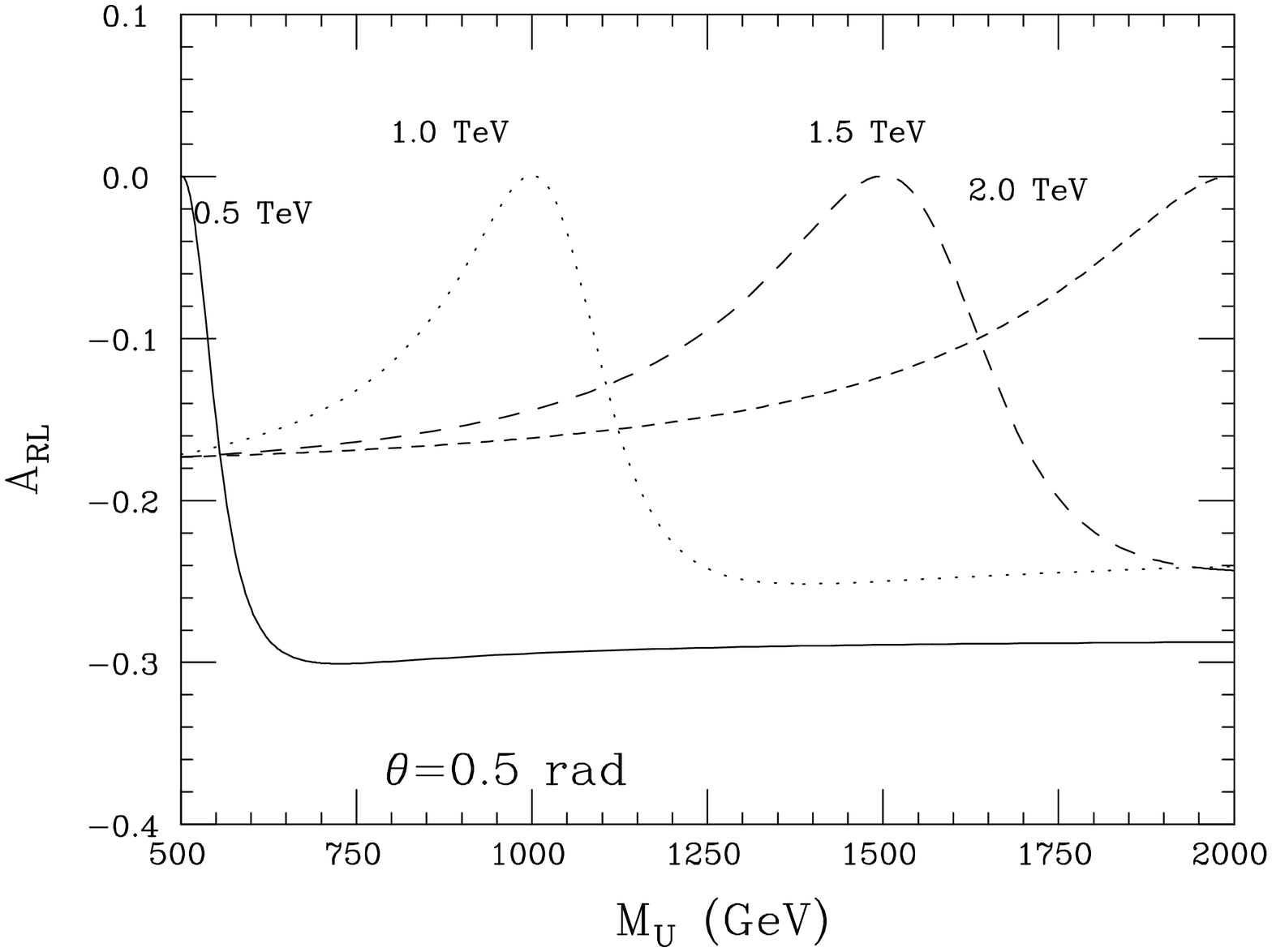,width=0.7\textwidth,angle=90}}       
\end{center}
\vglue 2cm
\caption{The $A_{RL}$ asymmetry  for a fixed scattering angle
$\theta=0.5$ rad and several values of $\sqrt{s}$ of $e^-e^-$ colliders for 
ESM+U as a function of $M_U$.}
\label{fig2}
\end{figure}

\begin{figure}[ht]
\begin{center}
\vglue -0.009cm
\mbox{\epsfig{file=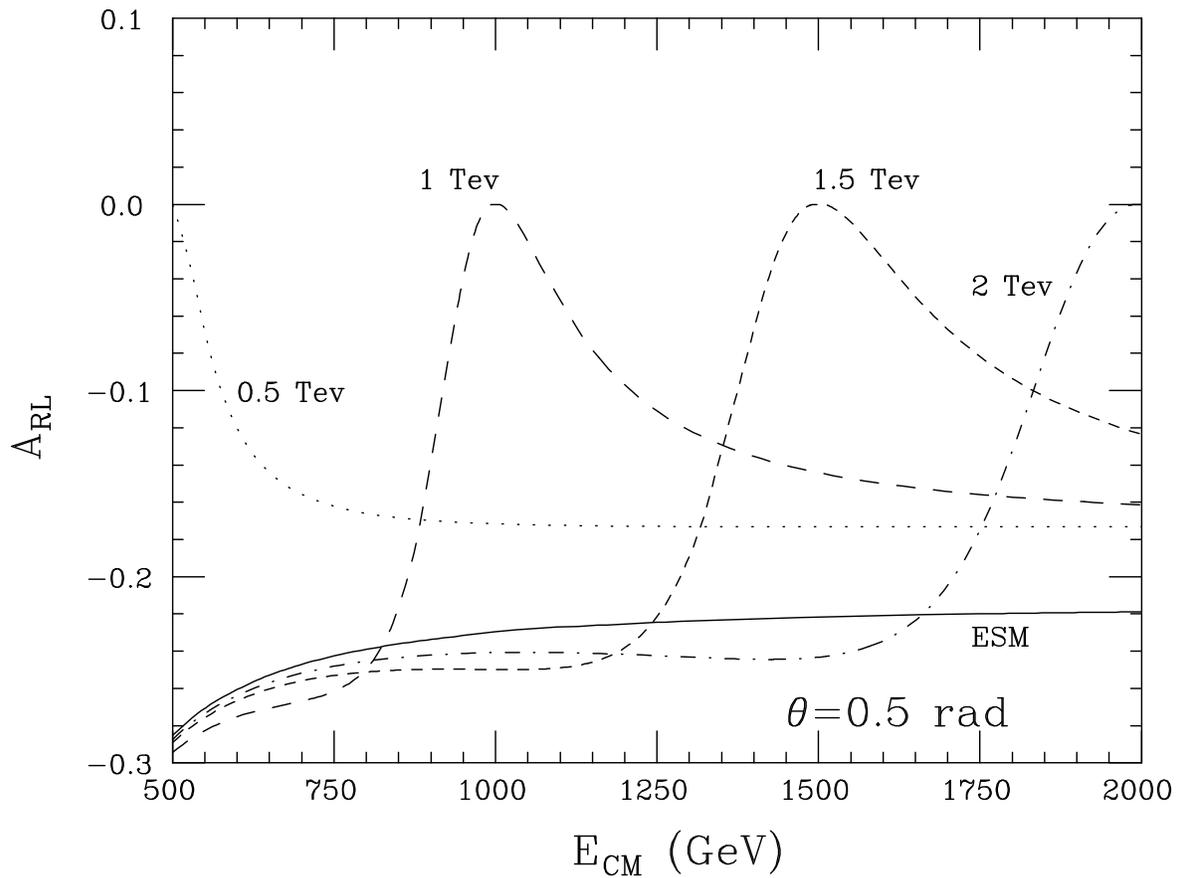,width=0.7\textwidth,angle=90}}       
\end{center}
\vglue 2cm
\caption{The $A_{RL}$ asymmetry  for a fixed scattering angle
$\theta=0.5$ rad for the ESM (solid line) and for the ESM+U for several 
U-masses as a function of $E_{CM}$.}
\label{fig3}
\end{figure}

We start this section by considering the contributions of the doubly charged
vector boson $U$ to the asymmetries which contributes uniquely via the
$s$-channel for a doubly charged initial state. 
In Fig.~\ref{fig1} we see that the angular 
dependence of the $A_{RL}$ asymmetry, taken into account the $U$ contribution, 
presents a relatively different behavior with respect to the ESM for a 
wide range of  $U$-masses. Notice that the lines are considerably 
separated even for those values of the $U$-mass that are not close to 
$\sqrt{s}$. Notice also that for $U$-mass lower than $\sqrt{s}$ we basically
reproduce the ESM result for $\theta\approx0$ and $\pi/2$; 
the largest difference with the ESM occurs for $\theta$ in the
interval 0.5--1.
The behavior of $A_{RL}$ for ESM+U as a function of $M_U$ is
showed in Fig.~\ref{fig2} for a fixed scattering angle and for several 
values of $\sqrt{s}$. We see that this asymmetry is essentially negative and
that its maximum value is zero and occurs at the resonance point
$M_U=\sqrt{s}$. This is due to the fact that at  the resonance the numerator 
of $A_{RL}$, as defined in Eq.~(\ref{asy1}), cancels out no mater the value of 
$M_U$. On the other hand the value of $M_U$ governs the width of the curves 
around the resonance point. This particular feature is better seen in
Fig.~\ref{fig3}. In this figure we show the $A_{RL}$ asymmetry as a function 
of the center-of-mass energy $E_{CM}=\sqrt{s}$ for some values of the 
$U$-mass and it is clearly seen that not only at the peak but also for a
considerably large range of masses around the peak, the curves representing 
the respective $U$ contribution are significantly separated from the ESM one.
It means that this asymmetry is  very sensitive to  the $U$-boson 
even in the case where the $U$-mass is larger than $\sqrt{s}$; when the 
$U$-mass is lower than $\sqrt{s}$ we reproduce the ESM results. 

\begin{figure}[ht]
\begin{center}
\vglue -0.009cm
\mbox{\epsfig{file=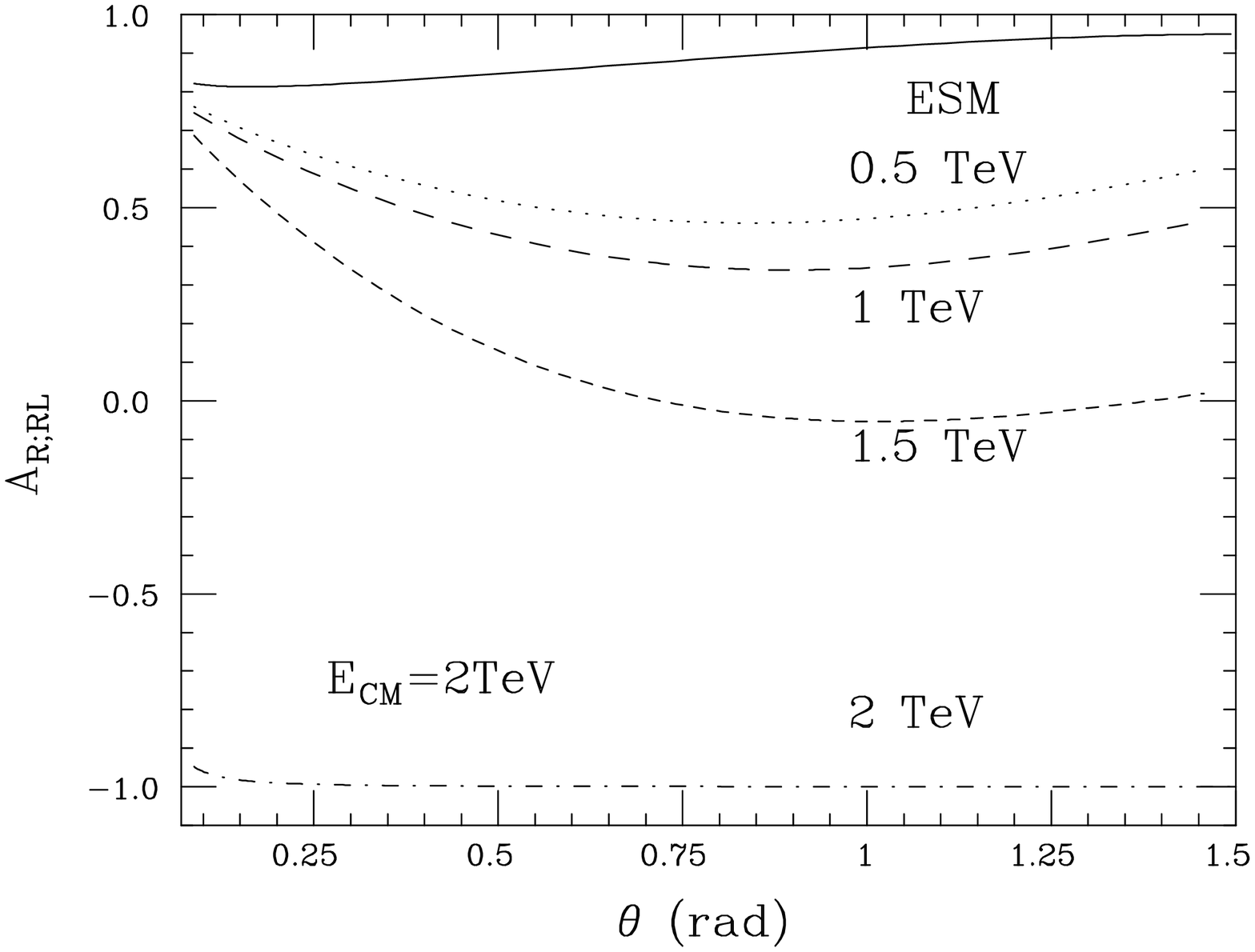,width=0.7\textwidth,angle=90}}       
\end{center}
\vglue 2cm
\caption{The same as in Fig.~1 for the $A_{R;RL}$ asymmetry.}
\label{fig4}
\end{figure}

\begin{figure}[ht]
\begin{center}
\vglue -0.009cm
\mbox{\epsfig{file=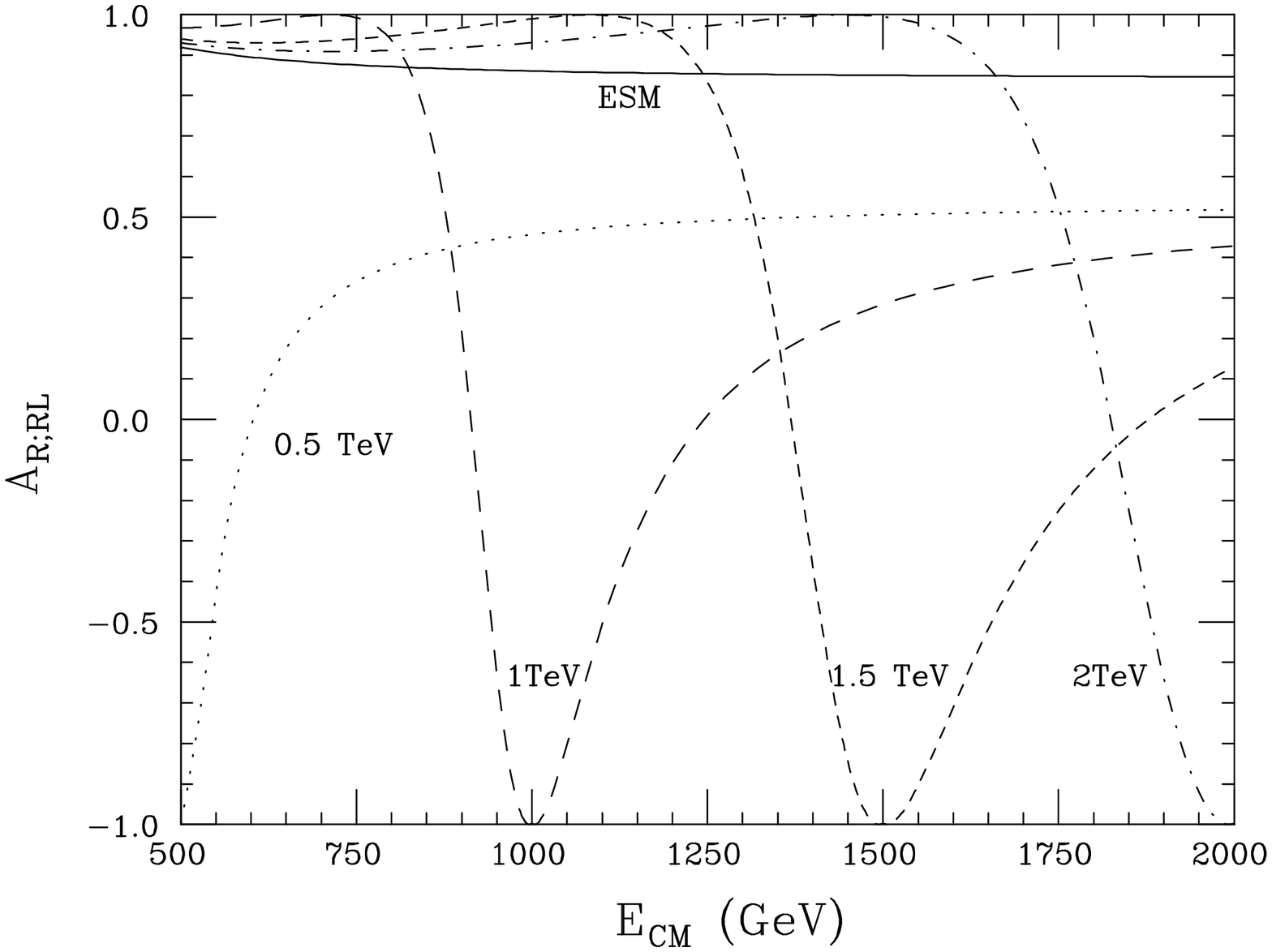,width=0.7\textwidth,angle=90}}       
\end{center}
\vglue 2cm
\caption{The same as in Fig.~3 for the $A_{R;RL}$ asymmetry .}
\label{fig5}
\end{figure}

In Fig.~\ref{fig4} and Fig.~\ref{fig5} we show the effects of the $U$-boson 
on the $A_{R;RL}$ asymmetry, defined in Eq.~(\ref{ar}),  and as it behaves
qualitatively like $A_{RL}$ we  come to the same conclusions we did for 
$A_{RL}$. We must note that near the $U$-resonance the $A_{R;RL}$ asymmetry
is negative. However, in this case, polarization for both beams 
must be available. 

\begin{figure}[ht]
\begin{center}
\vglue -0.009cm
\mbox{\epsfig{file=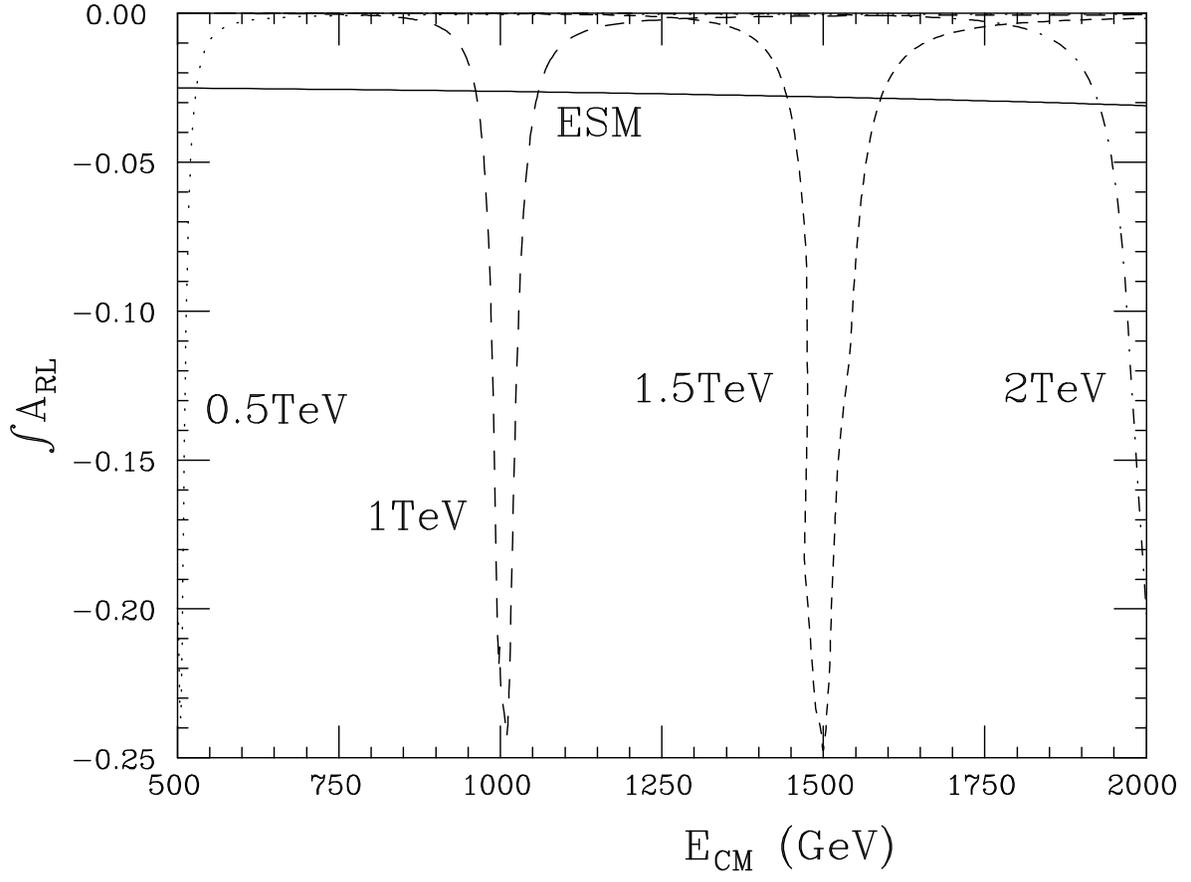,width=0.7\textwidth,angle=90}}       
\end{center}
\vglue 2cm
\caption{The  integrated asymmetry $\int A_{RL}$ for $e^-e^-$ collider 
for the ESM (solid line) and for the ESM+U for several U-masses as 
function of $E_{CM}$.}
\label{fig6}
\end{figure}

\begin{figure}[ht]
\begin{center}
\vglue -0.009cm
\mbox{\epsfig{file=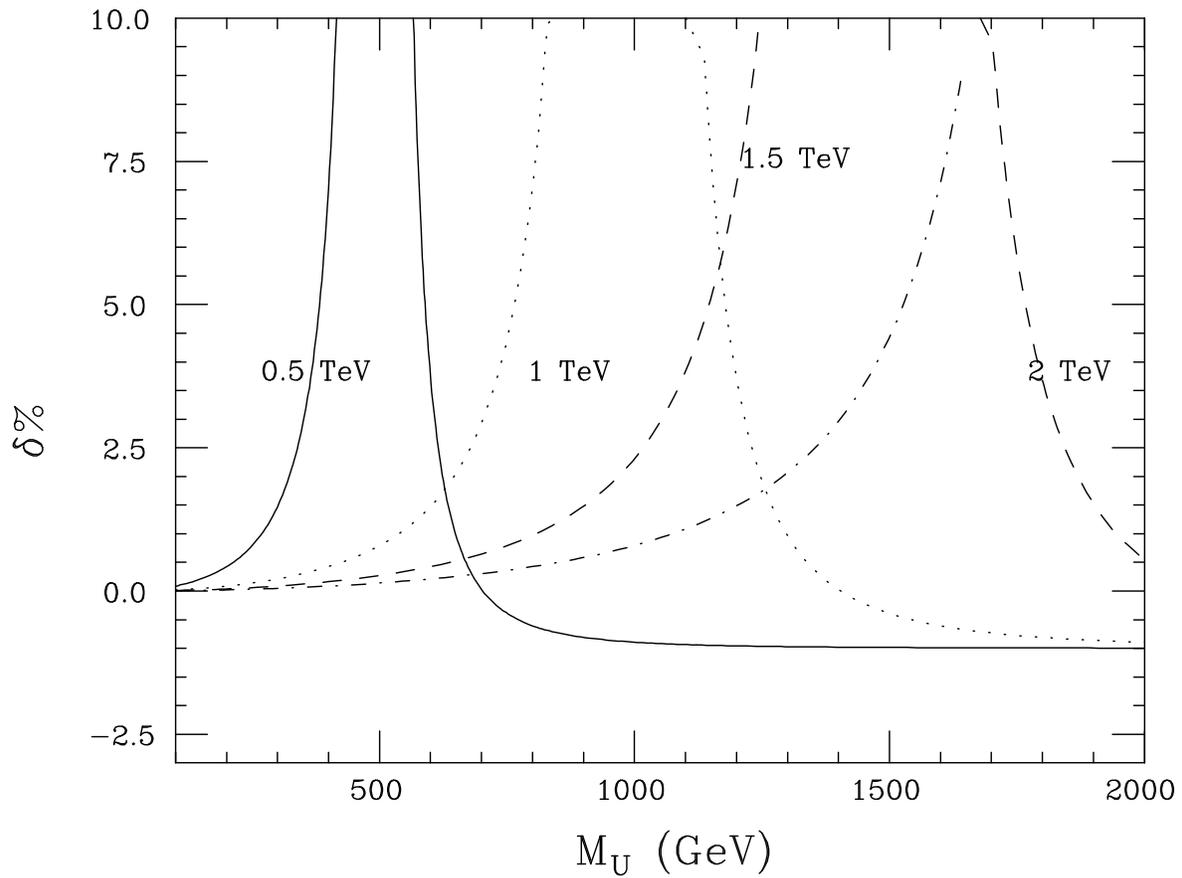,width=0.7\textwidth,angle=90}}       
\end{center}
\vglue 2cm
\caption{The quantity $\delta\,\%$ as defined in Eq.(\ref{delta}) for 
$e^-e^-$ collider for several
$\sqrt{s}$--values as a function of the $M_U$.}
\label{fig7}
\end{figure}

The integrated asymmetry $\int A_{RL}$ defined in Eq.~(\ref{aint}) is shown
in Fig.~\ref{fig6}. There we can see that while the ESM curve keeps an almost 
constant value (0.025-0.031) for $0.5 <\sqrt{s}< 2$ TeV, the ESM+U curves 
go from zero, for $M_U\not=\sqrt{s}$, to  a very pronounced 
peak ($\sim -0.25$) at the resonance points. In Fig.~\ref{fig7}, for the sake
of detectability, we show the quantity $\delta\%$ defined by:
\begin{equation}
\delta\,\% = \frac{\int A_{RL}^{ESM+U}-
                                   \int A_{RL}^{ESM}}
                                   {\int A_{RL}^{ESM}}\times 100,
\label{delta}
\end{equation}
which in this case stands for the percent deviation of 
$\int A_{RL}^{ESM+U}$ from $\int A_{RL}^{ESM}$. There we can see
that there is a  wide range of $U$-masses that can be probed at $e^-e^-$
colliders.         

\begin{figure}[ht]
\begin{center}
\vglue -0.009cm
\mbox{\epsfig{file=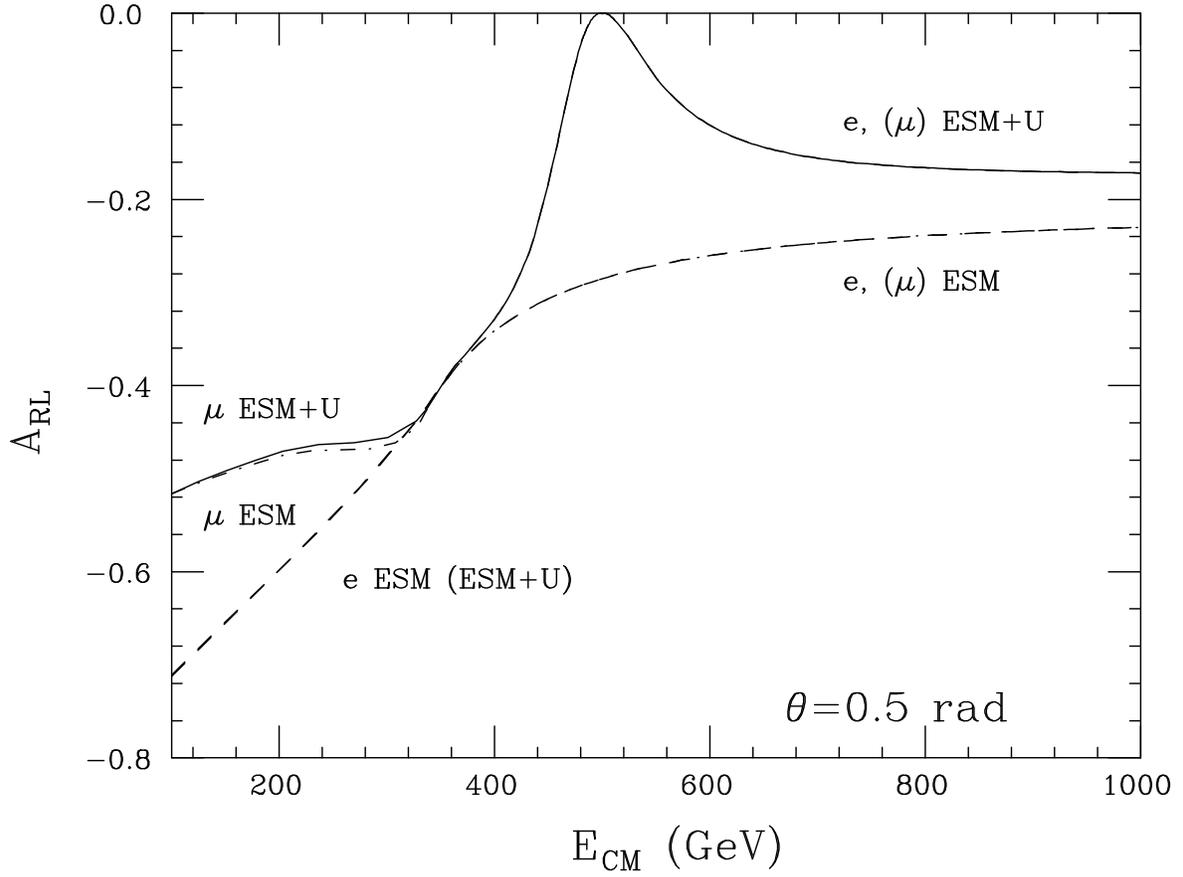,width=0.7\textwidth,angle=90}}       
\end{center}
\vglue 2cm
\caption{The $A_{RL}$ asymmetry for $M_U=0.5$ TeV and $\theta=0.5$ rad for
$e^-e^-$ and $\mu^-\mu^-$ colliders for the ESM and for the ESM+U as a function
of $E_{CM}$.}
\label{fig8}
\end{figure}

Next we study the effect of  non-negligible initial and final fermion 
masses by considering the $\mu^-\mu^- \to \mu^-\mu^-$ process in a muon 
collider. The results are given in Fig.~\ref{fig8}. There we can see that,
for $M_U=500$ GeV,  
below 300 GeV the muon mass effect is in evidence differing sensibly 
from the electron-electron case, independently of
the $U$-contribution for higher energies the lepton mass has no effect at all
for both models. Between 300 and 400 GeV all the curves are coincident and
we cannot distinguish among both models or leptons. 
Above 400 GeV it is the $U$-effect which dominates
and it is the same for electrons and muons. The effect of the $U$-resonance is
well evident and even above the resonance there is an almost constant
difference between the ESM+U and ESM asymmetries showing that the $A_{RL}$
asymmetry is still a sensitive parameter for the $U$ discovery for
$\sqrt{s}>M_U$ in lepton-lepton colliders.

\subsection{The $Z^\prime$ boson}                                   

\begin{figure}[ht]
\begin{center}
\vglue -0.009cm
\mbox{\epsfig{file=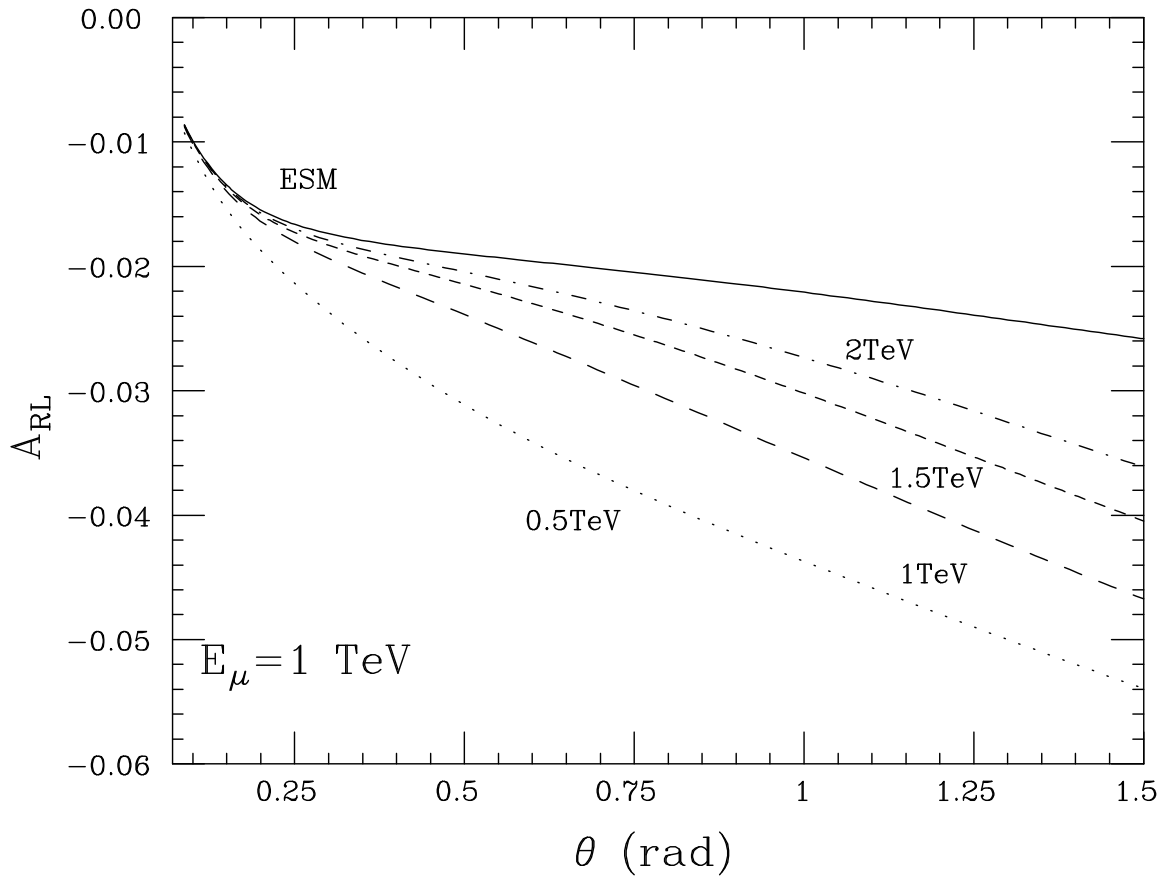,width=0.7\textwidth,angle=90}}       
\end{center}
\vglue 2cm
\caption{The $A_{RL}$ asymmetry for a $\mu^-e^-$--collider of $E_\mu=1$ Tev 
($\sqrt{s}=2$ TeV) for the ESM (solid line) and for the ESM+Z$^\prime$ for
several values of $M_{Z^\prime}$ as function of the scattering angle.}
\label{fig9}
\end{figure}

\begin{figure}[ht]
\begin{center}
\vglue -0.009cm
\mbox{\epsfig{file=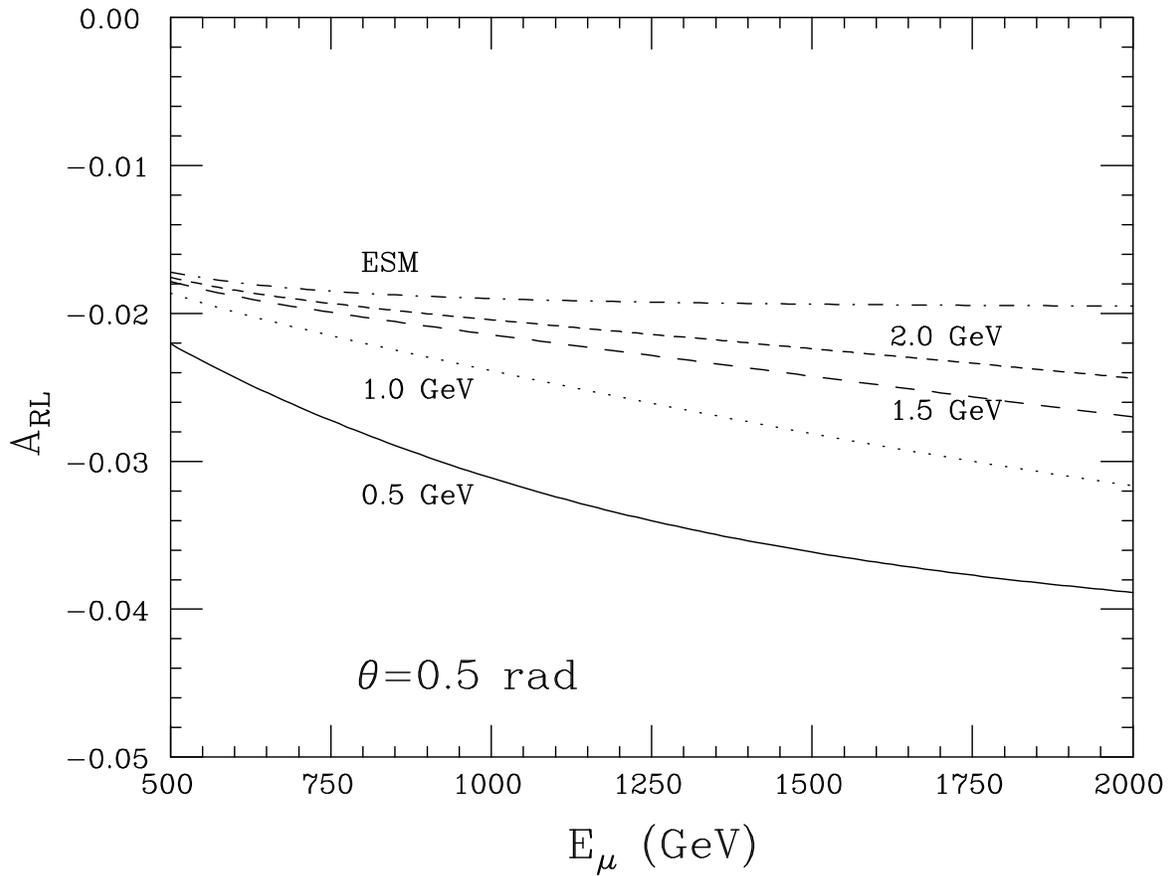,width=0.7\textwidth,angle=90}}       
\end{center}
\vglue 2cm
\caption{The $A_{RL}$ asymmetry for a $\mu^-e^-$--collider 
for a fixed scattering angle, $\theta=0.5$ rad, for the ESM (solid line) and for the ESM+Z$^\prime$ for
several values of $M_{Z^\prime}$ as function of $E_\mu$.}
\label{fig10}
\end{figure}

\begin{figure}[ht]
\begin{center}
\vglue -0.009cm
\mbox{\epsfig{file=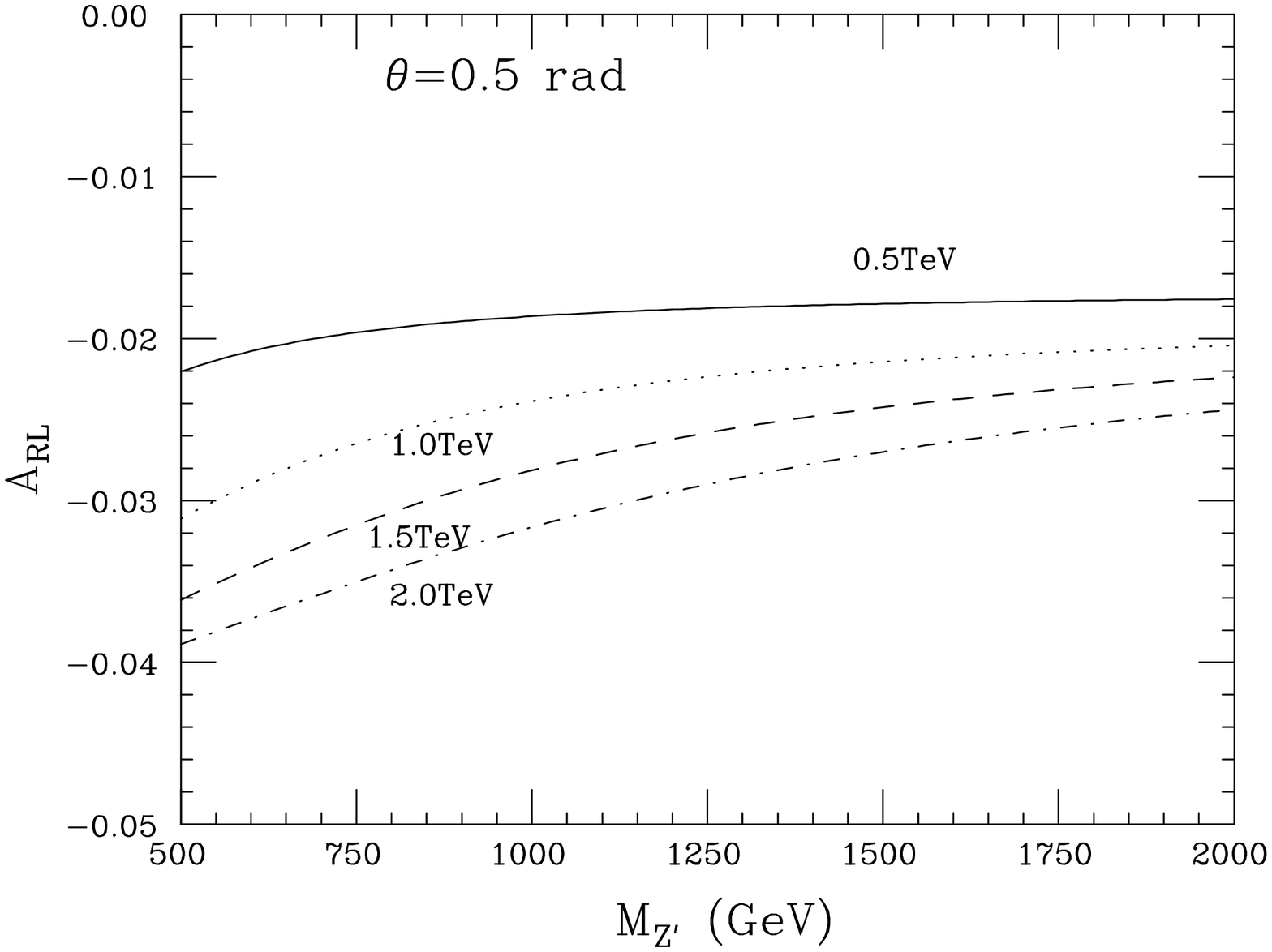,width=0.7\textwidth,angle=90}}       
\end{center}
\vglue 2cm
\caption{The $A_{RL}$ asymmetry  for a fixed scattering angle,
$\theta=0.5$ rad, and several values of $\sqrt{s}$ of $\mu^-e^-$ colliders for 
ESM+Z' as a function of $M_{Z^\prime}$.}
\label{fig11}
\end{figure}

In order to search for new physics in the neutral-vector boson sector it is
worth to consider de non-diagonal  process $\mu^-e^- \to  \mu^-e^-$. In this
case, assuming that the couplings of the $U$-boson with leptons are almost
diagonal ($K_{ii}\sim 1$ as in the $\Gamma_U$), the $s$-channel $U$-boson 
exchange will be negligible and provided that the $Z^\prime$ couples 
with leptons  diagonally, the only contributions to this process for 
ESM+Z$^\prime$ will be the $t$-channel ones, i.e., 
the contributions of $\gamma,Z$,  and $Z^\prime$. 
The angular dependence of $A_{RL}$ is showed in Fig.~\ref{fig9} where we 
can see that a  $Z^\prime$ contribution is clearly distinguished from the 
ESM one for a wide range of $Z^\prime$-masses around a given 
$E_\mu=\sqrt{s}/2$.
(In the figures concerning the $Z^\prime$-boson, once there is no $s$-channel, 
we specify the energy by the muon-beam energy $E_\mu$.)
As expected the $A_{RL}$ asymmetry is more sensitive to relatively light
$Z^\prime$ boson.
We come to the same conclusion from Fig.~\ref{fig10} in which we show the 
$A_{RL}$ asymmetry as a function of $E_\mu$. The sensitivity of 
the this asymmetry with the $Z^\prime$-mass is showed in Fig.~\ref{fig11}.

\begin{figure}[ht]
\begin{center}
\vglue -0.009cm
\mbox{\epsfig{file=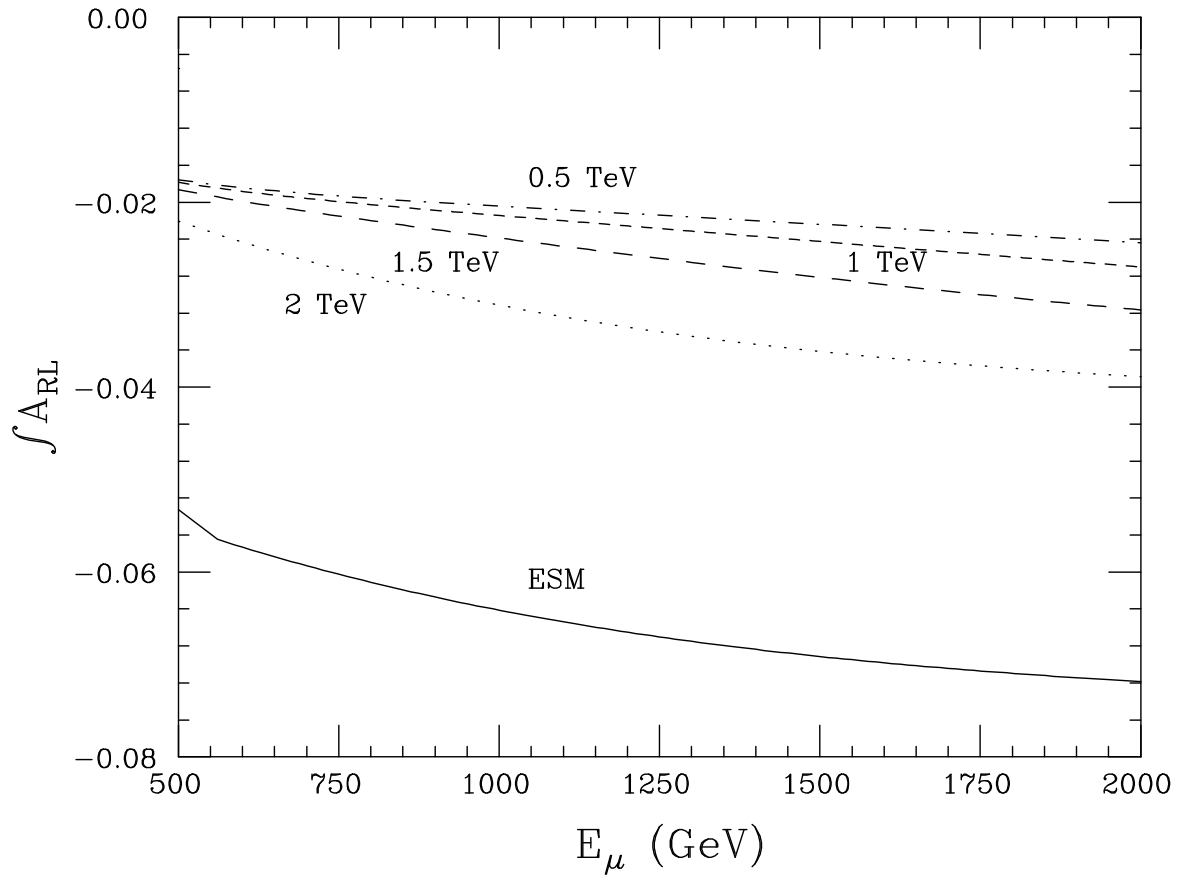,width=0.7\textwidth,angle=90}}       
\end{center}
\vglue 2cm
\caption{The integrated asymmetry $\int A_{RL}$ for the ESM (solid line) and
for the ESM+Z$^\prime$ for several $M_{Z^\prime}$  as function of $E_{\mu}$ 
for $\mu^-e^-$ colliders .}
\label{fig12}
\end{figure}

Contrarily to the case of the search for the $U$-boson, the asymmetry 
$A_{R;RL}$ is not sensitive to the extra neutral vector boson.
In this case, the potential capabilities of the  asymmetries in
discovering new neutral vector bosons are better explored by considering the 
integrated asymmetry  $\int A_{RL}$. The angular integration over the 
scattering angle of the $t$-channel  $Z^\prime$ exchange contribution produces 
curves that are clearly separated, depending on the $Z^\prime$-mass, which are 
also clearly distinguishable from the ESM curve for a wide range of masses. 
See Fig.~\ref{fig12}. 

We have also computed the asymmetries taking into
account a  $Z^\prime$ which couples to leptons with the same couplings of the 
standard Z boson but with a different mass. Although these ESM couplings are
stronger than those of the 3-3-1 model they have no substantial effect on 
the asymmetries: The results are very similar to the ones showed in 
Fig.~\ref{fig11}. 

For the $e\mu$ scattering there are also 
contributions coming from the neutral scalar sector of the model. 
However as in all the scalar contributions the pure scalar terms 
cancel out in the numerator of the asymmetry and the interference terms
are numerically negligible~\cite{assi1}. 

\subsection{Observability}
\label{subsec:ob}

Based on the figures we have shown throughout the text we have claimed that 
the values of the asymmetries, when there is an extra contribution of a 
new vector boson, are different enough from those of ESM ones to allow for 
the discovery of the referred bosons. However, we must be sure that there is 
enough statistics to measure these asymmetries. In order to provide some 
statistical analysis we assume a conservative value for the  
luminosity: 
${\cal L}=1\,\mbox{fb}^{-1}\mbox{yr}^{-1} = 10^{32}\mbox{cm}^{-2}\mbox{s}^{-1}$
for the $e^-e^-$, $\mu^-\mu^-$ and the $\mu^-e^-$ colliders, and compute the 
number of the expected number of events ($N$) based on the unpolarized 
integrated cross section for each process.  

For the $e^-e^- \to e^-e^-$ or $\mu^-\mu^-\to\mu^-\mu^-$ processes, the ESM 
cross section is relatively small, it goes from $0.05$ nb at $\sqrt{s}=0.5$ 
TeV to $1.5\times 10^{-3}$ nb at $\sqrt{s}=2$ TeV. These values correspond to 
$5\times 10^4$ and $1.5\times10^3$ events/yr respectively. 
Then, computing $\sqrt{N}/N$ we get $\sim 4\times 10^{-3}$, for the first case,
and $\sim 2\times 10^{-2}$,  for the second one.  This is an indication that
the ESM asymmetries can be measured. Note that the asymmetries we have computed
here are relatively large: 
$A_{RL}^{ESM}\sim A_{R;RL}^{ESM}\sim {\cal O}(10^{-1})$ 
for a fixed scattering angle and of the order 
${\cal O}(10^{-2})$ for the integrated ones as showed in 
Figs.~\ref{fig1}-\ref{fig5} and Fig.~\ref{fig6}, respectively. 
On the other hand, this reaction is so sensitive to the $U$-boson contribution 
that there is an enormous enlargement in the cross section and consequently in
the statistics. In Tab.~\ref{t1} we show the relevant parameters depending on
the $U$-boson mass and for only two values of $\sqrt{s}$ for shortness. 
There we can see that there is enough precision to measure
the asymmetries, for the range of masses and energies we have considered. 
For the $U$-boson discovery we can study the cross section directly 
provided that it is considerably different from that of the ESM~\cite{ph}. 
However, the
study of the asymmetry gives us more qualitative information once, 
contrarily to the cross section, it filters the vector-nature contribution of 
the $U$-boson: it was previously shown~\cite{assi1} that the scalar-boson 
contributions, which are present in the 3-3-1 model, cancel out in the 
asymmetry numerator. 

For the $\mu^-e^- \to \mu^-e^-$ process, the integrated cross sections for the
ESM are larger than for the electron diagonal process. We find $3$ nb for
$E_\mu=0.5$ TeV and $\sim  0.2$ nb for  $E_\mu=2.0$ TeV. The corresponding
number of events are $3\times 10^6$ and $\sim 2\times 10^5$, respectively, for
the same luminosity  used before. 
In this case the ratio $\sqrt{N}/N$ gives $5.7\times 10^{-4}$ and 
$2.2\times 10^{-3}$, respectively, which provide enough precision to measure 
the $A_{RL}^{ESM}$ once it is of the order ${\cal O}(10^{-2})$, as can be seen
from Fig.~\ref{fig9}-\ref{fig12}. For the ESM+Z$^\prime$ case we have that
although the Z$^\prime$-contribution can affect significantly 
the values of the asymmetry  $A_{RL}$, as it has been shown, it only slightly 
modifies the cross section values. It means that the number of events in for 
ESM+Z$^\prime$ are similar to those of the ESM and hence we have enough
precision to measure $A_{RL}^{ESM+Z^\prime}$ too. Once again we note that 
the asymmetries are more sensitive than the cross section itself
in looking for the Z$^\prime$-discovery.

\section{Conclusions}
\label{sec:con}

Here we have generalized the analysis of Ref.\cite{assi1,assi2} and have shown
that the L-R asymmetries in the diagonal ($e^-e^-, \mu^-\mu^-$) lepton
scattering can be the appropriate observable to discover doubly charged vector
bosons $U$ even for values of $M_U$ and $\sqrt{s}$ far away from the resonance
condition. Although the cross sections may have important contributions from
the scalar fields (doubly charged Higgs bosons) these contributions cancel out
in the numerator of the L-R asymmetries. On the other hand, the contribution of
the extra neutral Z$^\prime$, leptophobic or with the same couplings of the
standard model, gives small contributions to the diagonal $e^-e^-, \mu^-\mu^-$ 
scattering but it gives an important contribution to non-diagonal $\mu^-e^-$
case.
  
Hence, both $U^{--}$ and Z$^\prime$ vector bosons can be potentially
discovered in these sort of processes by measuring the L-R asymmetries. 
Since the couplings of both particles with matter are known in a given model,
once their masses were known other processes like exotic decays could be used
to study the respective contributions of the scalar fields present in the
model.

\acknowledgments 
This work was supported by Funda\c{c}\~ao de Amparo \`a Pesquisa
do Estado de S\~ao Paulo (FAPESP), Conselho Nacional de 
Ci\^encia e Tecnologia (CNPq) and by Programa de Apoio a
N\'ucleos de Excel\^encia (PRONEX).

\narrowtext

\vglue 0.01cm

\vglue 0.01cm

\vglue 0.01cm

\vglue 0.01cm

\vglue 0.01cm

\vglue 0.01cm

\vglue 0.01cm

\vglue 0.01cm

\vglue 0.01cm

\vglue 0.01cm

\vglue 0.01cm

%

\vglue 0.01cm

\newpage
\begin{table}
\caption{The quantities $\sigma$, $N$, and $\sqrt{N}/N$
 for the unpolarized $e^-e^- \to e^-e^-$ process with ${\cal L}=1\,
\mbox{fb}^{-1}\mbox{yr}^{-1}$ for the ESM and the ESM+U ($U$-mass dependent);
masses and energies are in TeV units.}
\begin{center}
\begin{tabular}{|l|c|c|c|c|c|l|} 
$\sqrt{s}=$ & \multicolumn{3}{c|}{$0.5$} &
      \multicolumn{3}{c|}{$2.0$}     \\ \hline
Model & $\sigma$ (nb)       & $N (\mbox{events})$ & $\sqrt{N}/N$  &
        $\sigma$ (nb)       & $N (\mbox{events})$ & $\sqrt{N}/N$ \\ \hline 
ESM   & $5\times 10^{-2}$   & $5\times 10^4$      & $4\times 10^{-3}$ &
        $1.5\times 10^{-3}$ & $1.5\times 10^3$    & $2\times 10^{-2}$ \\
         \hline \hline
ESM+U &  \multicolumn{6}{c|}{}     \\ \hline 

$M_U=0.5$  & $427$       & $4\times 10^{8}$    & $5\times 10^{-5}$   &
             $1.2$       & $1.2\times 10^{6}$  & $9\times 10^{-4}$  \\
                \hline  
$M_U=1.0$  & $25$        & $2.5\times 10^{7}$     & $2\times 10^{-4}$ &
             $\sim 1.2$       & $\sim 1.2\times 10^{6}$     & $\sim 9\times 10^{-4}$
             \\ \hline
$M_U=2.0$  & $\sim 25$   & $\sim 2.5\times 10^{7}$  & $\sim 2\times 10^{-4}$ & 
             $26$        & $ 2.6\times 10^{7}$      & $\sim 2\times 10^{-4}$
\end{tabular}
\end{center}
\label{t1}
\end{table}

\end{document}